\documentclass[preprint,12pt]{aastex}
\usepackage{epsfig}
\usepackage{emulateapj5}
\usepackage{apjfonts}
\usepackage{graphicx}

\newcommand{\chandra}{{\it Chandra}}

\newcommand{\rosat}{{\it ROSAT}}

\newcommand{\lum}{\thinspace\hbox{$\hbox{ergs}\thinspace\hbox{s}^{-1}$}}
\newcommand{\flux}{\thinspace\hbox{$\hbox{ergs}\thinspace\hbox{cm}^{-2}\thinspace\hbox{s}^{-1}$}}
\newcommand{\hst}{{\it HST}}

\makeatletter
\newenvironment{inlinetable}{%
\def\@captype{table}%
\noindent\begin{minipage}{0.999\linewidth}\begin{center}\footnotesize}
{\end{center}\end{minipage}\smallskip}

\newenvironment{inlinefigure}{%
\def\@captype{figure}%
\noindent\begin{minipage}{0.999\linewidth}\begin{center}}
{\end{center}\end{minipage}\smallskip}

\begin{document}

\def\spose#1{\hbox to 0pt{#1\hss}}
\def\laeq{\mathrel{\spose{\lower 3pt\hbox{$\mathchar"218$}}
     \raise 2.0pt\hbox{$\mathchar"13C$}}}
\def\gaeq{\mathrel{\spose{\lower 3pt\hbox{$\mathchar"218$}}
     \raise 2.0pt\hbox{$\mathchar"13E$}}}

\slugcomment{Accepted for publication in ApJ}

\title{{\it Chandra} and {\it Hubble Space Telescope} Study of the 
Globular Cluster NGC 288 $^1$} 

\author{Albert~K.~H.~Kong\altaffilmark{2}, 
Cees Bassa\altaffilmark{3}, David~Pooley\altaffilmark{4,7}, 
Walter~H.~G.~Lewin\altaffilmark{2}, Lee~Homer\altaffilmark{5}, 
Frank~Verbunt\altaffilmark{3},
Scott~F.~Anderson\altaffilmark{5} and Bruce Margon\altaffilmark{6}
} 
\altaffiltext{1}{Based on observations made with
the NASA/ESA Hubble Space
  Telescope, obtained at the Space Telescope Science Institute, which
  is operated by the Association of Universities for Research in
  Astronomy, Inc., under NASA contract NAS 5-26555. These observations
  are associated with program 10120 and 9959.}
\altaffiltext{2}{Kavli Institute for Astrophysics and Space Research,
Massachusetts Institute of Technology, 77
Massachusetts Avenue, Cambridge, MA 02139; akong@space.mit.edu}
\altaffiltext{3}{Astronomical Institute, Utrecht University, P.O. Box 
80000, 3508 TA, Utrecht, the Netherlands}
\altaffiltext{4}{Department of Astronomy, University of California, 601 
Campbell Hall, Berkeley, CA 94720-3411}
\altaffiltext{5}{Department of Astronomy, University of Washington, Box 
351580, Seattle, WA 98195}
\altaffiltext{6}{Space Telescope Science Institute, 3700 San Martin 
Drive,
  Baltimore, MD 21218}
\altaffiltext{7}{{\it Chandra} Fellow}

\begin{abstract}
We report on the {\it Chandra X-ray Observatory} observations of the 
globular cluster NGC 288. We detect four X-ray sources within the core 
radius and seven additional sources within the half-mass radius down 
to a limiting luminosity of $L_X=7\times10^{30}$\lum\ (assuming
cluster membership) in the 0.3--7 keV
band. We also observed 
the cluster with the {\it Hubble Space Telescope} Advanced Camera for
Surveys and identify 
optical counterparts to seven X-ray sources out of the nine sources
within the \hst\ field-of-view. Based on the X-ray and 
optical properties, we find 2--5 candidates of cataclysmic variables
(CVs) or chromospherically active
binaries, and 2--5
background galaxies inside the half-mass radius. Since the core
density of NGC 288 is very low, the faint X-ray sources of NGC 288 found in the \chandra\ and \hst\
       observations
is higher than the prediction on the basis of the collision frequency.
We suggest that the CVs and chromospherically active binaries are primordial in origin, in agreement
   with theoretical expectation.
\end{abstract}

\keywords{binaries: close---globular clusters: individual 
(NGC 288)---novae, cataclysmic variables---X-rays: binaries}

\section{Introduction}

The number of bright X-ray sources, with luminosities greater than
10$^{36}$ erg s$^{-1}$, per star has been estimated to be $\sim$100 times
as large in Galactic globular clusters as in the Galactic disk (Katz
1975; Clark 1975).
A population of dim sources, with X-ray luminosities less than about
$10^{34.5}$ erg s$^{-1}$, was discovered later (Hertz \&\ Grindlay 1983;
see also Verbunt 2001). The result by Pooley et al. (2003) suggests that the number of dim sources with
$L_{0.5-6 keV}>4\times 10^{30}$ erg s$^{-1}$ scales roughly with the
number of close stellar encounters in a cluster, and that the
incidence per star of dim X-ray sources is also higher in globular clusters than in the galactic disk.
There is evidence, from X-ray bursts
associated with the nuclear burning of accreted matter, that
the bright sources are neutron stars which are accreting matter
from a companion. Although
some of the dim sources may be neutron stars or even black holes
in quiescence, it is likely
that a majority of these, especially those with the lowest
luminosities, are accreting white dwarfs and X-ray active main
sequence/sub-giant binaries (see e.g., Pooley et al. 2003; 
Verbunt \& Lewin 2004; Heinke et al. 2005).

To identify faint X-ray sources in globular clusters, the \chandra\ X-ray
Observatory and
{\it Hubble Space Telescope (HST)} are absolutely essential because the density of X-ray sources 
is expected to be high and sub-arcsecond spatial resolution is
required to search for reliable optical counterparts in such crowded regions. Since the launch
of \chandra, four globular clusters dominated by faint X-ray sources have been studied in detail with
\chandra\ and \hst: 47 Tuc (Grindlay et al. 2001a; Heinke et al. 2005), NGC 6397 (Grindlay
et al. 2001b), NGC 6752 (Pooley et
al. 2002a), and M4 (Bassa et al. 2004). Of these faint X-ray sources,
many are believed to be cataclysmic variables (CVs) and X-ray active
binaries (e.g. RS CVn and BY Dra systems). There are also quiescent
low-mass X-ray binaries (qLMXBs; Heinke et al. 2003a and references therein) and millisecond
pulsars (Edmonds et al. 2001; Grindlay et al, 2001b; Bassa et al. 2004).

We report here new \chandra\ and \hst\
observations of the globular cluster NGC 288. NGC 288 is  
a globular cluster
 with a fairly low central density (log $\rho_0=1.8 L_{\odot}$ pc$^{-3}$; 
Djorgovski 1993) with a core radius of $85''$ and a 
half-mass radius of $2.25'$ (Trager, Djorgovski, \& King 1993). The 
distance to this cluster is estimated at 8.4 kpc (Peterson 1993). 
The reddening is quite low with $E(B-V)=0.03$, corresponding to
a neutral hydrogen column $N_H=1.6\times10^{20}$ cm$^2$ (Predehl \&\
Schmitt 1995). NGC 288 is located close to the Southern Galactic Pole ($l=152.28^{\circ}, b=-89.38^{\circ}$) and thus we are looking
directly out of the plane. As such, there will not be many foreground
objects at optical and X-ray wavelengths. 
The cluster 
center is given by Webbink (1985) as R.A.=00$^h$52$^m$45.3$^s$ and 
Decl.=$-26^{\circ}34'43''$ (J2000). The absolute visual magnitude of
NGC 288 is
$-6.7$ (Harris 1996, version of February 2003). An exceptionally high concentration of 
blue stragglers and binary systems in the core of NGC 288 suggests
that the blue
stragglers' production mechanism via binary evolution can be very 
efficient (Bolte 1992; Bellazzini et al. 2002). In the X-ray waveband, NGC 288 has 
only been observed with \rosat\ HRI (Sarazin et al. 1999).
From the \rosat\ data, only one 
X-ray source was within the half-mass radius (Sarazin et al. 1999).

In \S2, we describe our \chandra\ observations and analysis of NGC 288. 
We discuss the \hst\ observations in \S3, and source identification in 
\S4. A discussion and comparison with other globular clusters will be 
given in \S5.

\begin{inlinefigure}
\medskip
\psfig{file=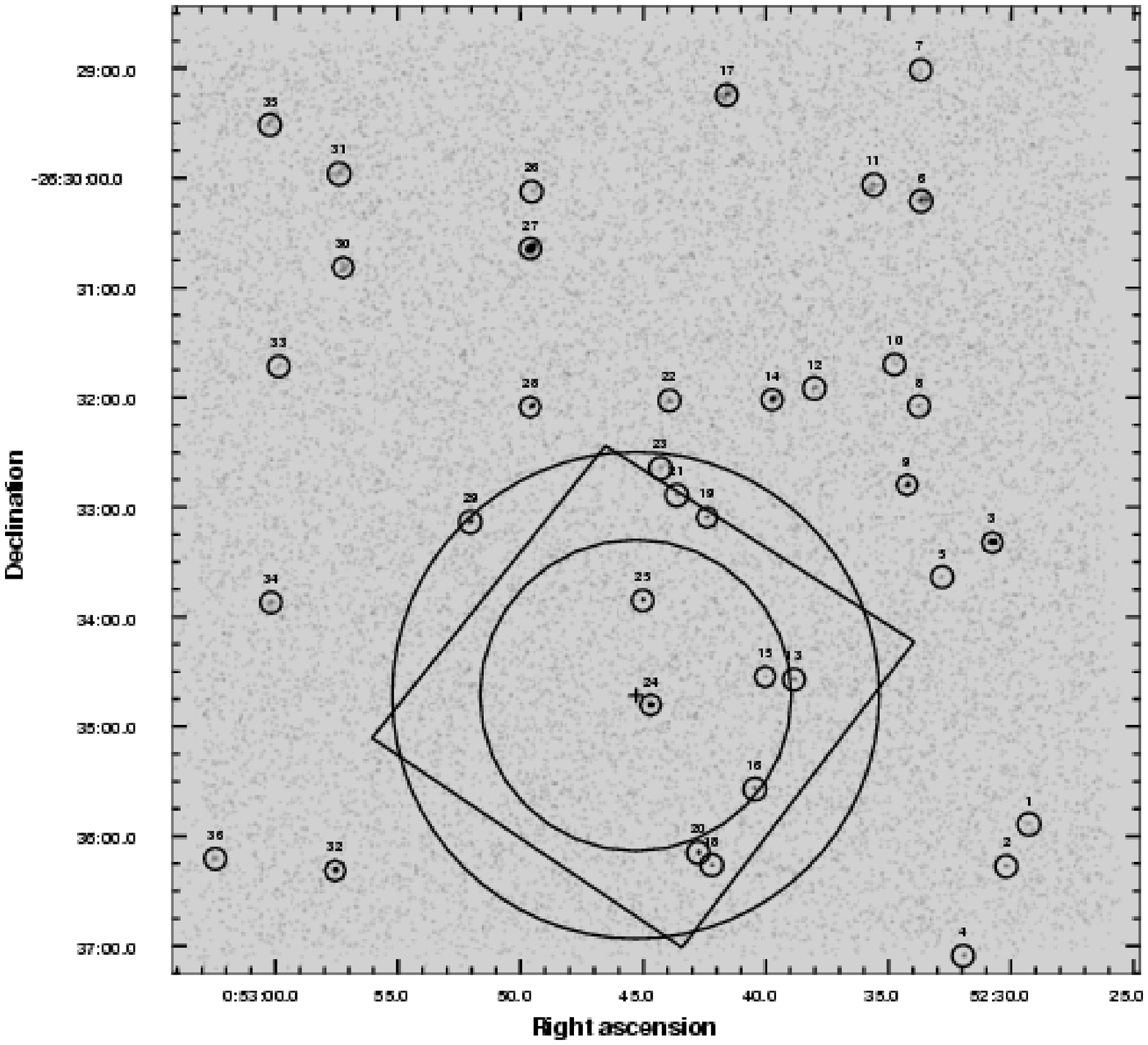,width=3.6in}
\caption{\chandra\ ACIS-S3 0.3--7 keV image of NGC 288. The image was
smoothed with a Gaussian function. The large circle
represents the half-mass radius and the inner circle is the core radius.
The cluster center is marked by a cross. The square is the field-of-view
of the \hst/ACS. The detected X-ray sources are marked and numbered.
}
\end{inlinefigure}

\section{X-ray Observations and Analysis}
NGC 288 was observed with \chandra\ for 55 ks on 2003 Feburary 6 (OBSID 
3777). The telescope aim point is on the Advanced CCD 
Imaging Spectrometer (ACIS) back-illuminated S3 chip. Data  
were telemetered in faint mode and were collected with a
frame transfer time of 3.2 s. The $\sim 8\times8$ arcmin$^2$ S3 chip 
covers the entire cluster half-mass radius. In this paper, we only focus 
on the data taken with the S3 chip.

\subsection{Data Reduction}
The data reduction and analysis was done with CIAO,
Version 3.2.2 and ACIS Extract\footnote{http://www.astro.psu.edu/xray/docs/TARA/ae\_users\_guide.html} 
(Broos et al. 2002).
We reprocessed the level 1 event files with CALDB 3.1.0.
In order to reduce the instrumental background, only data with {\it 
ASCA} grades of 0, 2, 3, 4, and 6 were
included. Only events with
photon energies in the range of 0.3--7.0 keV were included in our
analysis. We also inspected the
background count rates from the S1 chip; about 6 ks was eliminated due 
to high background count rate (count rate $> 2$ counts s$^{-1}$). The effective exposure time for the 
observation after filtering for background flares was 49 ks. 

\subsection{Source Detection}

Discrete sources in the \chandra\ images were found with WAVDETECT 
(Freeman et
al. 2002) together with exposure maps. We performed source detection on 
the 0.3--1 keV, 1--2 keV, 2--7 keV, and 
0.3--7 keV images. We set the detection threshold to be $10^{-6}$, 
corresponding to less than one false detection due to statistical 
fluctuations in the background. For each of the four energy band
images, we performed source detection 
using sequences of wavelet scales that increased by a factor of 
$\sqrt{2}$ from scales 1 to 16.
We then combined the four source 
lists into a master source list. 
A total of 36 X-ray sources were
detected. Figure 1 shows the detected X-ray sources on the ACIS-S3 chip. 
Source counts in the three band passes were extracted 
from polygonal extraction 
regions which approximate 90\% of the \chandra\ point spread 
function (PSF).
Background was extracted from a circle
centered on each source but excluding the 99\% contour of the PSF 
around any point source.

Table 1 lists the 36 \chandra\ sources in our catalog, sorted in order 
of increasing right ascension. The
columns give the source number, the position (J2000.0), the net counts
in the three energy bands (soft: 0.3--1 keV; medium: 1--2 keV;  
hard: 2--7 keV), 
the 0.5--2.5 and the 0.3--7 keV unabsorbed fluxes. The conversion to fluxes assumes an 
absorbed power-law spectrum with a photon index of
2 and $N_H = 1.6\times10^{20}$ cm$^{-2}$. If we assume a thermal
bremsstrahlung model with $kT=10$ keV, then the flux will
be $\sim 10$\% higher than that of power-law model. It is worth noting
that CX 15 is a very soft X-ray source with all photons coming from
the 0.3--1 keV band. If we assumed  a power-law model and that the counts were from the whole 0.3--7 keV range, the flux would be underestimated. Therefore, for CX 15, we convert the
flux by assuming counts from 0.3--1 keV only. For the power-law
model, the resulting 0.3--7 keV unabsorbed  flux is $1.5\times10^{-15}$\flux. If we
assume a blackbody model with $kT=0.1$ keV, then the flux will
be $7.8\times10^{-14}$\flux. 
The detection limit in the
0.5--2 keV band of our 
observation is about $3.2\times10^{-16}$ ergs cm$^{-2}$ s$^{-1}$.
We estimated the number of 
background sources using \chandra\ deep field
data (Brandt et al. 2001). Using the log $N$--log $S$ distribution
derived from the deep field data, 
between 26--36 sources out of the 36 sources are background objects in 
the ACIS-S3. Within the half-mass radius, there are 11 sources and 7--9 
sources are estimated to be background. For an expected number of 9 background sources, the probability of
  finding 11 or more background sources is 30\%. Thus we cannot
  exclude that all our sources are background sources. Indeed, in the
  \chandra\ image shown in Figure 1 the surface number density is not
  noticeably higher within the halfmass radius than outside it.

\begin{table*}
\centering{\footnotesize
\caption{\chandra\ Source Properties}
\begin{tabular}{lccccccccc}
\hline
\hline
\multicolumn{2}{c}{Source} & R.A. & Decl. &\multicolumn{3}{c}{Net Counts} & $f_{0.5-2.5}$& $f_{0.3-7}$ & \\
\cline{1-2} \cline{5-7}
Name & CXOU J & (J2000.0) & (J2000.0) & Soft & Medium & Hard & &  &
Counterpart 
\\
\hline
CX 1 & 005229.4-263553 & 00:52:29.413 (0.30) & -26:35:53.46 (0.23) & 0.7 & 4.8 & 0.5 & 0.46 &0.91&\\
CX 2 & 005230.3-263616 &00:52:30.336 (0.21) & -26:36:16.52 (0.23) &  1.7 & 2.8 & 2.6 & 0.54 &1.07&\\
CX 3 &005230.9-263319 & 00:52:30.909 (0.07) & -26:33:19.47 (0.04) & 98.7 & 69.8 & 28.5 & 15.16&29.85&\\
CX 4 &005232.0-263705 & 00:52:32.092 (0.12) & -26:37:05.36 (0.18) &  0.9 & 11.0 & 5.9 & 1.37&2.7&\\
CX 5 &005232.9-263338 & 00:52:32.950 (0.22) & -26:33:38.28 (0.09) &  1.8 & 2.8 & 2.6 & 0.56&1.09&\\
CX 6 &005233.8-263012 & 00:52:33.818 (0.24) & -26:30:12.67 (0.23) & 22.7 & 27.2 & 7.4 & 4.43&8.68&\\
CX 7 &005233.8-262901 & 00:52:33.838 (0.26) & -26:29:01.05 (0.68) &  4.8 & 5.9 & 5.5 & 1.26&2.45&\\
CX 8 &005233.8-263204 & 00:52:33.899 (0.28) & -26:32:04.80 (0.21) & 4.5 & 0.0 & 5.3 & 0.76&1.48&\\
CX 9&005234.3-263247 & 00:52:34.380 (0.09) & -26:32:47.90 (0.09) & 12.7 & 12.8 & 9.6 & 2.71&5.31&\\
CX 10&005234.8-263142 & 00:52:34.897 (0.39) & -26:31:42.08 (0.25) &  2.4 & 1.6 & 4.1 & 0.33&1.22&\\
CX 11&005235.7-263003 & 00:52:35.748 (0.40) & -26:30:03.80 (0.14) &  15.7& 7.2 & 1.1 & 1.86&3.64&\\
CX 12&005238.1-263155 & 00:52:38.162 (0.19) & -26:31:55.14 (0.11) & 1.6 & 0.0 & 13.3 &1.14& 2.25&\\
CX 13&005238.9-263434 & 00:52:38.991 (0.13) & -26:34:34.29 (0.09) & 0.0 & 0.0 & 7.7 & 0.60&1.17& \hst \\
CX 14&005239.8-263201 & 00:52:39.888 (0.11) & -26:32:01.33 (0.08) &  9.7 & 15.8& 13.5 & 3.02&5.91&\\
CX 15&005240.1-263432 & 00:52:40.164 (0.23) & -26:34:32.82 (0.17) & 3.9 & 0.0 & 0.0 &0.32\tablenotemark{a}, 0.78\tablenotemark{b}&0.78$^a$, 1.54\tablenotemark{b}& \hst\\
CX 16&005240.5-263534 & 00:52:40.585 (0.18) & -26:35:34.32 (0.13) &  2.9 & 4.9 & 4.8 & 0.97&1.91&\\
CX 17&005241.7-262914 &  00:52:41.735 (0.23) & -26:29:14.74 (0.26) &  25.4 & 25.3& 13.9 & 5.01&9.83&\\
CX 18&005242.3-263615 & 00:52:42.337 (0.10) & -26:36:15.83 (0.15) & 0.0 & 6.9 & 3.8 & 0.83&1.62& \hst \\
CX 19&005242.5-263305 & 00:52:42.526 (0.14) & -26:33:05.71 (0.13) & 2.9 & 8.9 & 2.8 & 1.12&2.21& \hst \\
CX 20&005242.6-263609 & 00:52:42.622 (0.08) & -26:36:09.08 (0.08) & 8.9 & 8.9 & 7.8 & 1.97&3.89& \hst \\
CX 21&005243.7-263253 & 00:52:43.764 (0.12) & -26:32:53.54 (0.12) & 3.9 & 6.9 & 0.0 & 0.83&1.63&\\
CX 22&005244.0-263201 & 00:52:44.070 (0.22) & -26:32:01.78 (0.16) & 8.7 & 3.8 & 1.5 & 1.08&2.12&\\
CX 23&005244.4-263238 &  00:52:44.435 (0.18) & -26:32:38.88 (0.11) & 1.9 & 1.9 & 1.7 & 0.43&0.83&\\
CX 24&005244.8-263448 & 00:52:44.832 (0.03) & -26:34:48.22 (0.03) &
29.9 & 99.9 & 71.8 & 15.64&30.56& \hst, {\it ROSAT}\\
CX 25&005245.1-263351 & 00:52:45.158 (0.10) & -26:33:51.08 (0.06) & 11.9 & 15.9 & 7.8 & 2.76&5.40& \hst \\
CX 26&005249.6-263007 & 00:52:49.682 (0.25) & -26:30:07.47 (0.44) & 0.0 & 1.4 & 11.2 & 0.97&1.91&\\
CX 27&005249.7-263038 & 00:52:49.731 (0.06) & -26:30:38.63 (0.05) & 184.2 & 210.5 & 102.9 & 38.50&75.45&\\
CX 28&005249.7-263205 & 00:52:49.744 (0.11) & -26:32:05.28 (0.09) & 2.7 & 16.8 & 19.5 & 3.02&5.91&\\
CX 29&005252.1-263308 & 00:52:52.184 (0.10) & -26:33:08.26 (0.12) &  0.0 & 6.9 & 16.6 & 1.81&3.55&\\
CX 30& 005257.3-263048 &00:52:57.360 (0.18) & -26:30:48.93 (0.20) & 2.2 & 11.4 & 12.6 & 2.03&3.97&\\
CX 31&005257.5-262957 & 00:52:57.519 (0.28) & -26:29:57.76 (0.24) &  10.5 & 11.8 & 3.8 & 2.01&3.96&\\
CX 32&005257.7-263618 & 00:52:57.700 (0.08) & -26:36:18.90 (0.11)  &  42.8 & 5.9 & 0.6 & 3.80&7.47&\\
CX 33&005259.9-263143 & 00:52:59.980 (0.34) & -26:31:43.10 (0.27) & 2.4 & 2.5 & 3.5 & 0.65&1.26&\\
CX 34&005300.3-263352 & 00:53:00.307 (0.19) & -26:33:52.12 (0.15) & 0.0 & 5.7 & 7.4 & 1.02&1.99&\\
CX 35&005300.3-262931 & 00:53:00.341 (0.43) & -26:29:31.06 (0.45) &  2.1 & 11.7 & 12.4 & 2.01&3.97&\\
CX 36&005302.5-263612 & 00:53:02.583 (0.33) & -26:36:12.27 (0.20) &  0.7 & 5.8 & 3.4 & 0.77&1.50&\\
\hline
\end{tabular}
}
\par
\medskip
\begin{minipage}{0.95\linewidth}
NOTES. --- 
The positions in the table have been corrected for boresight, in that 
   the right ascensions and declinations resulting from the Chandra 
   source detection with {\it wavdetect} have been corrected by $0.137''$ and $-0.055''$
   respectively (see Sect. 3.2).
The positional uncertainties are in units of arcsec given
by {\it wavdetect}. The unabsorbed flux is in units of $10^{-15}$\flux\
and is derived assuming a power-law model (except for CX 15) with
$N_H=1.6\times10^{20}$ cm$^{-2}$ and a photon index of 2.\\
$^a$ Assuming a blackbody model with $0.1$ keV and counts from 0.3--1 keV.\\
$^b$ Assuming a power-law model with photon index of 2 and counts from 0.3--1 keV.
\end{minipage}
\end{table*} 

\subsection{X-ray Colors and Spectral Fitting}
Many of the sources in our catalog have $< 100$ counts, which makes it 
difficult to derive spectral parameters with meaningful constraints. 
However, hardness ratios can give a crude indication of the X-ray 
spectra in these cases. We therefore computed the hardness ratios for 
all the detected sources. These ratios were based on the source counts 
in three energy bands: S (0.3--1.0 keV), M (1--2 keV), and H (2--7 keV). 
The two hardness ratios are defined as HR1=(M-S)/(M+S) and HR2 = 
(H-S)/(H+S). Figure 2 shows the color-color diagram (left) and the
color-magnitude diagram (right) of all X-ray 
sources detected in the ACIS-S3 chip. We have overlaid the color-color 
diagram with four 
lines showing the tracks followed by representative spectra with 
differing values of $N_H$. Note that the colors in the
color-magnitude diagram were chosen to be consistent with previous work
(e.g. Pooley et al. 2002b; Heinke et al. 2003a; Bassa et al. 2004).

\begin{figure*}
\psfig{file=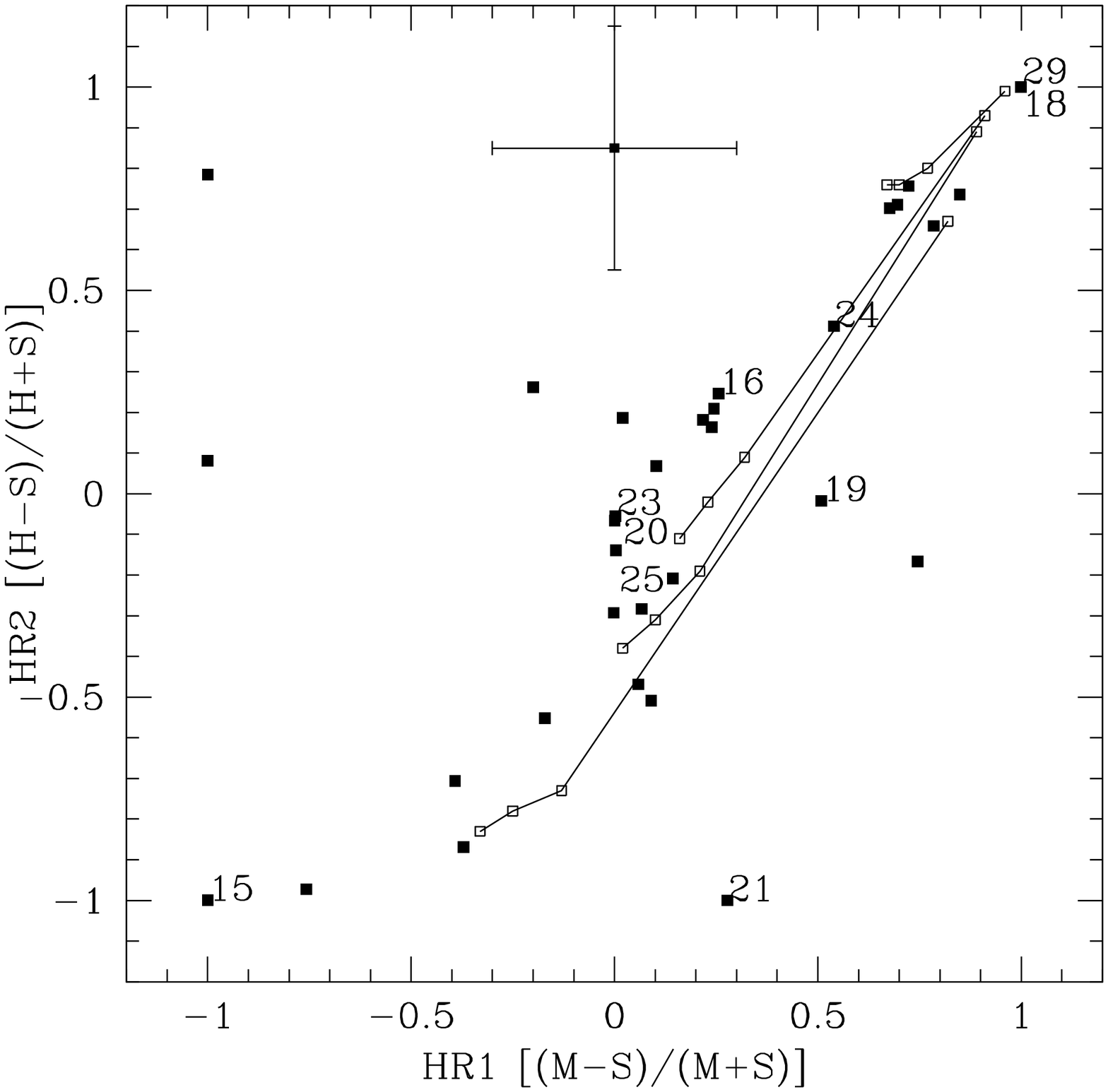,width=3.5in}
\psfig{file=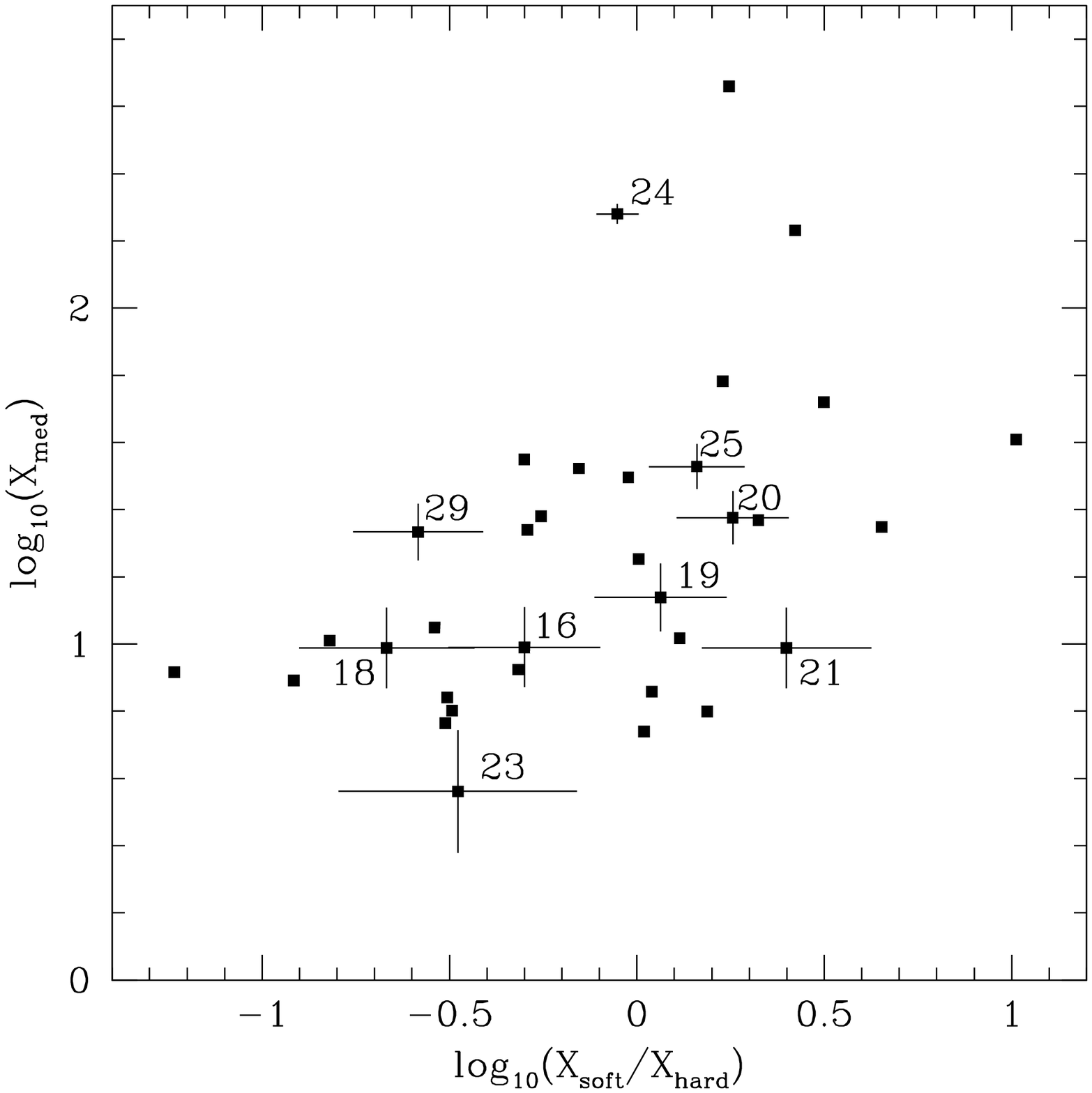,width=3.5in}
\caption{Left: X-ray color-color diagram of \chandra\ sources in NGC 288. The
  number corresponds to the source number inside the half-mass radius.
 Also 
plotted are the hardness ratios estimated from different 
spectral models. Top to bottom: Blackbody model with $kT=1$ keV, 
thermal bremsstrahlung model with $kT=5$ keV, power-law model with 
$\alpha=2$, and 3. For each model, $N_H$ (open squares) varies from the left as 
$1.6\times10^{20}$, $5\times10^{20}$, $10^{21}$, and $10^{22}$ cm$^{-2}$.
Right: X-ray color-magnitude diagram of NGC 288. The X-ray color is
  defined as the logarithm of the ratio of 0.5--1.5 keV (X$_{soft}$) counts to
  1.5--6 keV ($X_{hard}$) counts, and the magnitude is the logarithm of 0.5--4.5 keV
  ($X_{med}$) counts. Sources within the half-mass radius are marked
  with error bars.
}
\end{figure*} 

We extracted the energy spectra for the brightest three X-ray sources 
which have $\gaeq 200$ counts and fitted them to simple one-component 
spectral models including absorbed power-law and thermal bremsstrahlung 
models.  In 
order to employ $\chi^2$ statistics, all spectra were grouped into at 
least 15 counts per spectral bin. We forced $N_H$ to be $\geq 
1.6\times10^{20}$ cm$^{-2}$, the cluster $N_H$ derived from optical 
studies. All spectra can be 
fitted equally well with both models. Table 2 summarizes the spectral 
fits. For CX 3, the $N_H$ converged to values much lower than the
cluster $N_H$ derived from optical studies. For the other two sources, the
$N_H$ of the thermal bremsstrahlung model is slightly higher than the
cluster value, while it is much higher for power-law model. The
temperatures of the
thermal bremsstrahlung model vary between 1.7 and 7 keV, while the
photon index of the power-law model is consistent with 2.

\section{Optical Observations}
NGC 288 was observed with \hst\ Advanced Camera for Surveys (ACS) on 2004 
September 20 (Proposal ID: 10120). The 
observations consist of deep F435W ($B_{435}$), F625W ($r_{625}$), 
and F658N (H$\alpha_{658}$) images covering the core of the cluster. 
The exposure time 
with the F435W, F625W, and F658N filters is 740s, 320s, and 1760s, 
respectively. The ACS field-of-view 
covers the entire core radius of the cluster and about 73\% of the 
half-mass radius (see Figure 1). Three X-ray sources (CX16, 18 and 20) coincide with an archival \hst\
Wide Field and Planetary Camera 2 (WFPC2) observation of NGC\,288. For
this observations, the F255W ($\mathrm{nUV}_{255}$), F336W
($U_{336}$), F555W ($V_{555}$) and F814 ($I_{814}$) filters were
used. Exposure times were 700\,s in F255W, 3760\,s in F336W, 430\,s in
F555W and 585\,s in F814W.

\subsection{Data Reductions and Photometry}
Images of each bandpass were shifted and co-added
using the MultiDrizzle package in PyRAF, with masking of
cosmic rays, saturated pixels, and bad pixels. We used the combined 
images for correcting astrometry and identifying optical counterparts of 
X-ray sources in the cluster. Figure 3 shows the color \hst\ ACS
image of NGC 288. However, we used individual images to 
perform PSF photometry with the DOLPHOT package that is adapted from
HSTphot (Dolphin 2000) for the use of ACS data \footnote{http://purcell.as.arizona.edu/$\sim$andy/dolphot/}.  DOLPHOT 
is a stand-alone package to perform PSF photometry with a module for ACS 
data. We did not use the combined images for photometry because  
drizzled images require re-sampling producing suboptimal 
photometry. DOLPHOT can be run on multiple images of the same field and 
outputs the combined photometry for each filter. We first applied {\em 
acsmask} to mask bad pixels according to the data quality image provided 
by STScI. We then used {\em calcsky} to create sky images for 
background determination. Finally, we performed PSF photometry using 
DOLPHOT with lookup tables for the ACS PSF and produced a master list of positions and magnitudes for 
each star found. The final magnitudes were corrected for aperture and
charge transfer efficiency effects.
Additional selection criteria were applied to 
eliminate cosmic rays, artifacts, and ``stars'' lying on the 
diffraction spikes of the very brightest stars. The final photometry 
data were used to construct the color-magnitude diagrams (CMDs) shown in 
Figure 4. Stars are shown if they appear in all three filters.

The archival \hst/WFPC2 observation was photometered using HSTphot
1.1.5b (Dolphin 2000, see Bassa et al. 2004 for a more detailed
description). 

\subsection{Astrometry}

To identify optical counterparts to the \chandra\ X-ray sources in the 
field, we have to improve the astrometry of both datasets. 
We retrieved a 5-minute $V$-band image of NGC 288 with the Wide Field
Imager (WFI) at the ESO 2.2 meter telescope on La Silla, taken on 2004
June 14 and used that to calibrate the \hst/ACS images. The WFI has an array of 8 CCDs, each CCD having a $8'\times16'$ field of view, giving a total of $33'\times34'$. 
An $8'\times8'$ subsection of the WFI chip covering the cluster
center was used that
contained 93 UCAC2 standards (Zacharias et al. 2004). Of these, 72 were not saturated and
appeared stellar and unblended. Fitting for a 6 parameter
transformation, we obtained a solution giving residuals of $0.056''$ in 
R.A. and $0.059''$ in Decl.

The astrometry of the WFI image was then transferred to the two
ACS/WFC chips (WFC1 and WFC2). We used DOLPHOT to generate positions and
photometry for all stars on the ACS/WFC chips. These positions were
corrected for the considerable geometric distortion using
polynomials (Hack \& Cox 2001). A selection
of the stars on each chip (having $r_{625} <$18.0) were matched against
stars on the WFI image, where we used the distortion corrected
positions of ACS/WFC for comparison with the calibrated position on
the WFI. For WFC1, 174 stars were selected and outliers were removed
through an iterative process. The astrometric solution converged using
147 stars, yielding residuals of $0.016''$ in R.A. and $0.018''$ in 
Decl.  For
WFC2, we started off with 206 stars, while the final solution has 155
stars with residuals of $0.014''$ in R.A. and $0.016''$ in Decl. 

For the astrometric calibration of the archival \hst/WFPC2 images, we
first corrected all pixel positions of the stars for distortion and
placed them on a common frame using the distortion corrections and
relative chip positions and offsets by Anderson \& King (2003). The
resulting positions were matched to those of stars on the WFI image,
where the final astrometric solution used 207 stars giving residuals
of 0.022" in RA and 0.024" in Decl. 

\begin{inlinetable}
\caption{Spectral Fits of the Brightest Sources}
\begin{tabular}{lccccc}
\hline \hline
CX & Model\tablenotemark{a} & $N_H$\tablenotemark{b} & $kT/\alpha$ & $\chi^2_{\nu}/dof$ & $f_{0.3-7}$\tablenotemark{c}\\
\hline
3 & TB & $1.6^{+3.4}_{-0.0}$& $1.7^{+0.9}_{-0.5}$ & 1.3/10& 22.0\\
  & PL & $1.6^{+6.4}_{-0.0}$& $2.1^{+0.6}_{-0.2}$& 1.1/10& 28.0\\
24& TB & $3.0^{+1.4}_{-1.1}$ & $7.0^{+21.3}_{-3.4}$ & 0.9/10&  48.0\\
  & PL & $35^{+1.9}_{-1.1}$ & $1.7^{+0.4}_{-0.3}$& 0.9/10 & 36.0\\
27& TB & $3.9^{+3.1}_{-2.2}$& $4.4^{+2.4}_{-1.4}$& 1.0/28& 82.0\\
  & PL & $8.4^{+4.0}_{-4.0}$& $1.9^{+0.2}_{-0.2}$& 0.9/28 & 96.0 \\
\hline
\end{tabular}
\par
\medskip
\begin{minipage}{0.82\linewidth}
NOTES. --- All quoted uncertainties are 90\%.\\
$^a$ TB: thermal bremsstrahlung; PL: power-law.\\
$^b$ in units of $10^{20}$ cm$^{-2}$\\
$^c$ 0.3--7 keV unabsorbed flux in units of $10^{-15}$\flux.
\end{minipage}
\end{inlinetable}

For the \chandra\ image, we first used the Aspect 
Calculator\footnote{http://cxc.harvard.edu/ciao/threads/arcsec\_correction/} 
provided by the \chandra\ X-ray Center to correct the aspect offset. 
This will provide an absolute astrometry of $0.6''$ (90\%). The shifts 
were small: $0.07''$ in right ascension and $0.03''$ in declination.
Inspection of the 99\% confidence error circles of the \chandra\ X-ray
sources (using the positions from Table 1) on the WFI image yields
several likely counterparts.  X-ray sources CX 3, CX 12, CX 28, CX 30
appear to coincide with stellar objects, whereas CX 7, CX 8, CX 17, 
CX 22,
CX 26, CX 32, CX 36 appear to coincide with extended objects, possibly
background galaxies. 
Based on brightness, positional accuracy, and conformity of the image to the
   point spread function CX 28 and CX 30 are the most promising stellar
   counterparts. In addition, CX 28 and CX 30 are outside the
   half-mass radius, hence the stellar density from the cluster is
   relatively low implying a low probability of chance coincidences.
Based on these two optical counterparts, the boresight correction that
needs to be applied to the X-ray source positions is $0.137\pm0.104''$ 
in R.A. and $-0.055\pm0.096''$ in Decl.

\vspace{2mm}
\begin{inlinefigure}
\psfig{file=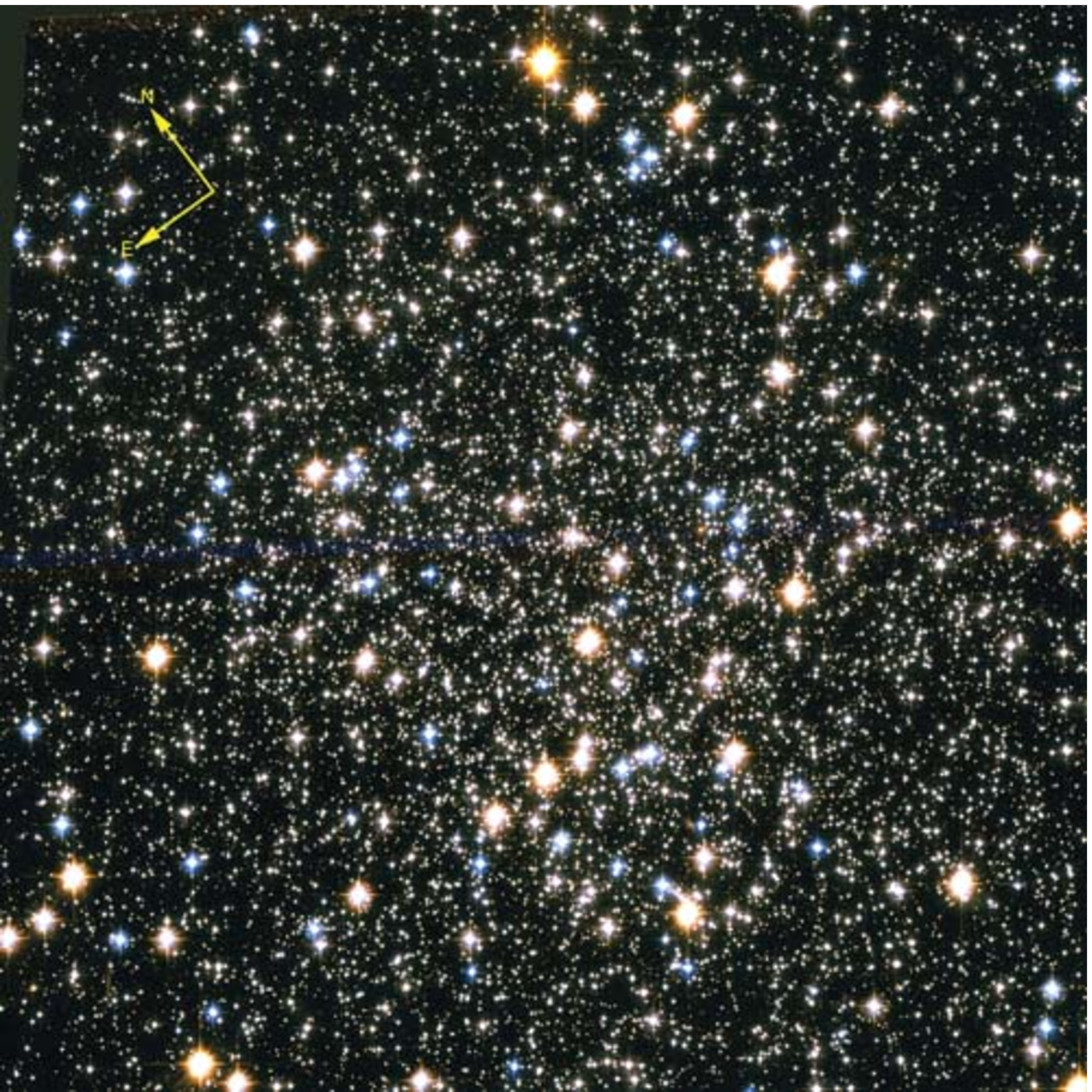,width=3.4in}
\caption{True color \hst\ ACS image ($202\times202$ arcsec) of NGC 288 core. This was constructed by
  combining all the $r_{625}$ (red), $0.5(B_{435}+r_{625})$ (green), and
  $B_{435}$ (blue) images.
}
\end{inlinefigure} 


\begin{table*}
\centering{\footnotesize
\caption{Optical Counterparts to \chandra\ X-ray Sources}
\begin{tabular}{lcccccccccc}
\hline
\hline
 & $\Delta$ R.A. & $\Delta$ Decl. & & & \\
CX & (arcsec) & (arcsec) & $U_{336}$ &$B_{435}$ & $V_{555}$
&$r_{625}$& $I_{814}$& H$\alpha_{658}$& $f_X/f_r$\tablenotemark{a} & Classification\tablenotemark{b}\\
\hline
13a & 0.09& -0.31& &$24.96\pm0.04$ && $22.90\pm0.02$ && $22.58\pm0.05$
& 0.8 & AB?\\
13b & 0.42& -0.08& &$26.04\pm0.09$ && $22.86\pm0.02$ && $22.34\pm0.04$
& 0.76 & CV? AGN?\\
15 & -0.27& 0.19& & $26.81\pm0.17$ && $25.78\pm0.19$ && $25.15\pm0.74$
& 11.6 & AGN\\
18 & -0.10& -0.26& $22.71\pm0.08$&$22.73\pm0.01$ &
$21.79\pm0.02$&$21.52\pm0.01$ & $20.78\pm0.02$&$21.16\pm0.02$ & 0.3 & AB\\
19a & 0.31& 0.007& &$25.24\pm 0.05$&& $25.33\pm0.11$ && $23.86\pm0.12$
& 13.7 & AGN\\
19b & -0.55& -0.21&& $20.54\pm0.004$& & $19.73\pm0.004$ &&
$19.50\pm0.013$ & 0.08 & AB\\
20a & -0.1& 0.09&& $27.51\pm0.29$ && $26.37\pm0.27$ && $24.78\pm0.42$
& 60.1 & AGN\\
20b & 0.16& -0.03& &$25.42\pm0.06$ & $24.34\pm0.15$& $24.19\pm0.05$ &
$22.88\pm0.08$&$23.74\pm0.12$ & 8.6 & CV\\
24 & -0.17& 0.007&& $24.88\pm0.04$ && $23.58\pm0.04$ && $23.47\pm0.10$
& 39.7 & CV\\
25 & 0.30& -0.05&& $26.92\pm0.19$ && $25.70\pm0.23$ && $24.05\pm0.19$
& 48.1 & AGN\\
\hline
\end{tabular}
}
\par
\medskip
\begin{minipage}{0.97\linewidth}
Note --- The last column give a tentative classification; for the
  sources with two possible optical counterparts, this classification
  holds only for the actual counterpart.\\
$^a$ Ratio of X-ray to optical ($r_{625}$) flux, using $log
  (f_X/f_r)=log f_X + 5.67 + 0.4 r_{625}$ (Green et al. 2004); $f_X$
  is derived in the 0.3--7 keV band.\\
$^c$ CV: cataclysmic variable; AB chromospherically active binary;
  AGN: active galactic nuclei
\end{minipage}
\end{table*}

\section{Source Identification and Classification}
To obtain optical identifications for the X-ray sources, we use the
precise astrometry described in the previous sections.
We searched for optical
counterparts within the 95\% \chandra\ error 
circle which is the quadratic sum of the positional uncertainty for the
X-ray source, the uncertainty in the optical astrometry (UCAC2 to WFI
astrometry and WFI to \hst/ACS astrometry), and the
uncertainty in the X-ray boresight correction. Within the ACS field-of-view, there are nine \chandra\ sources 
and we suggest optical counterparts based on positional coincidence alone
to seven of them. In case of multiple sources inside the error circle,
we included all the candidates within the 95\% X-ray error circle.
The results of each 
candidate optical counterpart are summarized in Table 3, and finding 
charts are shown in Figure 5. Using the photometric data from the \hst\ 
ACS, we constructed CMDs, shown in 
Figure 4.

To help in assessing the nature of the optically identified sources,
we show in Figure 6 the X-ray luminosity as a function of the absolute
magnitude, for low-luminosity X-ray sources in globular clusters. The
large symbols in this figure indicate the X-ray
sources with possible optical counterparts, in the field of view of our \chandra\ observation of NGC 288,
the smaller symbols show objects found in other clusters, mostly 47\,Tuc
and M4 (see Bassa et al.\ 2004). We note that the absolute magnitudes and
X-ray luminosities for the sources in our NGC 288 observations 
are computed under the assumption, which we will test below, that they 
are associated with NGC 288. As discussed earlier, we caution that the \chandra\ deep field data imply that for the 11 X-ray sources within the half-mass radius, there is
a 30\% probability that {\em all} of them are background sources.

We first consider the X-ray sources with only one suggested
counterpart in the error circle.
The ratio of the X-ray to optical flux locates CX 18 in the region 
of active binaries in Figure\,6, albeit close to the boundary with
CVs. The star in
the error circle of CX18 is located on the main-sequence in the
CMD of Figure 4. Because it
does not show noticeable H$\alpha$ emission, is on the main sequence
in $U_{336}-V_{555}$, and is not
detected in the near ultraviolet, it is unlikely to be a CV, and we
suggest that CX 18 is a chromosphically active binary. 


The candidate cluster counterpart to CX 24 is blue with respect to the main sequence, and has no H$\alpha$
emission, being located rather on the main sequence
in the H$\alpha$ diagram (Fig. 4). It
has a high X-ray to optical flux ratio and a hard spectrum with 0.3--7 keV
luminosity of $\sim 4\times10^{32}$\lum. It is worth noting that CX 24 is
the only X-ray source in the half-mass radius detected with \rosat\
(Sarazin et al. 1999). Using the \chandra\ spectral fit, the X-ray
luminosity during the \rosat\ observations is about $10^{33}$\lum, a
factor of 2.5 higher than our \chandra\ observation. We suggest that
CX 24 is a CV, even though its H$\alpha$ emission does not appear to be strong.

The sources CX 15 and CX 25 are optically extended, and thus almost certainly
background galaxies. Indeed, if we compute their X-ray luminosity and
optical magnitude under the wrong assumption that they are in
NGC\,288, we find that they are located in the $L_x-M_V$ diagram in an
area where no genuine cluster sources have been found. A probable
background quasar in the \chandra\ field of view of our M\,4 observation
is in the same location of Figure\,6 (for the wrong assumption
that it belongs to the cluster M\,4; Bassa et al. 2004, 2005; Bedin et
al. 2003).

Turning now to the sources with more than one possible counterpart in
the error circle, we first note that our suggested classifications
depend on the optical object indeed being the counterpart.  One star
in the error circle of CX 13, CX 13a, is on the main sequence both in
the CMD and in the H$\alpha$-$r$ diagram.  Like
CX 18, it may be an active binary. The other star, CX 13b is redder than
the main sequence. Its X-ray to optical flux ratio is somewhat high
for an active binary. Possibly it is an (obscured?) active
galaxy. Alternatively, its red color could be an artifact of variation
of its blue magnitude, in which case it may be a CV. (Note that we
have no evidence for such variation.)  In the
absence of more information, a secure classification of CX 13 thus is
not possible. 

Of the two objects in the error circle of CX 19, the brighter one (CX 19b)
would a priori be the more probable counterpart, since the probability
of a chance coincidence is higher for the more numerous faint objects.
Its colors (Figure\,4) and X-ray to optical flux ratio (Figure\,5)
then suggests that CX 19b is a chromospherically active binary. If the fainter object
CX 19a is the counterpart, its X-ray to optical flux ratio (Figure\,5)
suggests that it is a background galaxy/AGN. 

The brighter object in the error circle of CX 20, CX 20b is blue, and
has a relatively high X-ray to optical flux ratio (Figure\,6): it may
be a CV. CX 20b was also imaged by  WFPC2 
with the $V_{555}$ and $I_{814}$ filters. It has $V_{555}=24.34$ and is on the
main-sequence in $V-I$ (similar to CVs in other
 clusters, e.g. NGC6397 and 47 Tuc; Cool et al. 1998, Edmonds et al. 2003).
On the other hand, CX 20a is extended,
and therefore an galaxy, an active galaxy when it is the counterpart
of CX 20. CX 20a is below the detection limit
in the archival WFPC2 observations.

There are two unidentified X-ray sources (CX 16 and CX 21) in the ACS
field-of-view. CX 16 is in between two bright stars and the spikes
produced by these bright stars prevent us from searching for any faint optical
sources inside the \chandra\ error circle. The region of CX 16 was also
observed with the WFPC2, but neither do we find a counterpart inside the
\chandra\ error circle in these data.
CX 21 is near the edge of
the field and part of the \chandra\ error circle is in the dithering
pattern of the image. Therefore the sensitivity is greatly
reduced. Two additional sources (CX 23 and CX 29) are inside the
half-mass radius but were not observed with \hst.
In any case, if the identifications of CX 18 and CX 24 are correct, then these two remaining sources are very likely part of the extra-galactic
  background.

In summary, we find one good candidate CV
(CX 24) and the source is already detected with ROSAT (Sarazin et al.\ 1999),
and one good candidate active binary (CX 18).  The X-ray to optical
luminosity ratio of CX 24 is in the upper range of the values observed
for CVs in globular clusters so far, and the X-ray
to optical luminosity ratio of CX 18 is amongst the highest observed so
far for active binaries in globular clusters.
In addition to these, we have possible cluster members and according
classifications in the error circles of CX 13 (active binary or CV),
CX 19 (active binary) and CX 20 (CV). 

\begin{figure*}
\psfig{file=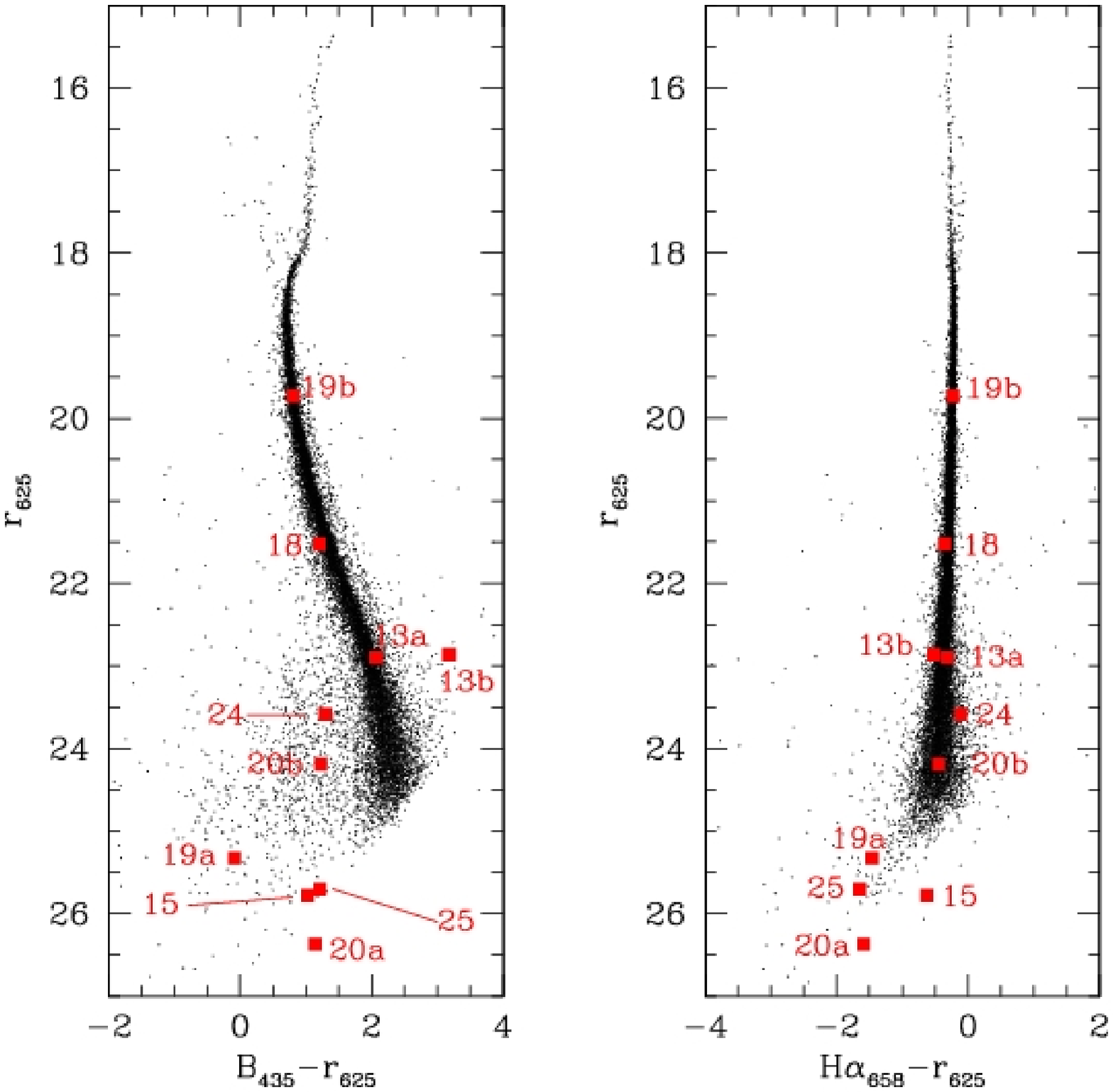,width=7.5in}
\caption{Color-magnitude diagram of the \hst\ ACS observations of 
NGC 288. The numbers refers to the candidate optical counterparts to the 
X-ray sources.
}
\end{figure*}

\begin{figure*}
\psfig{file=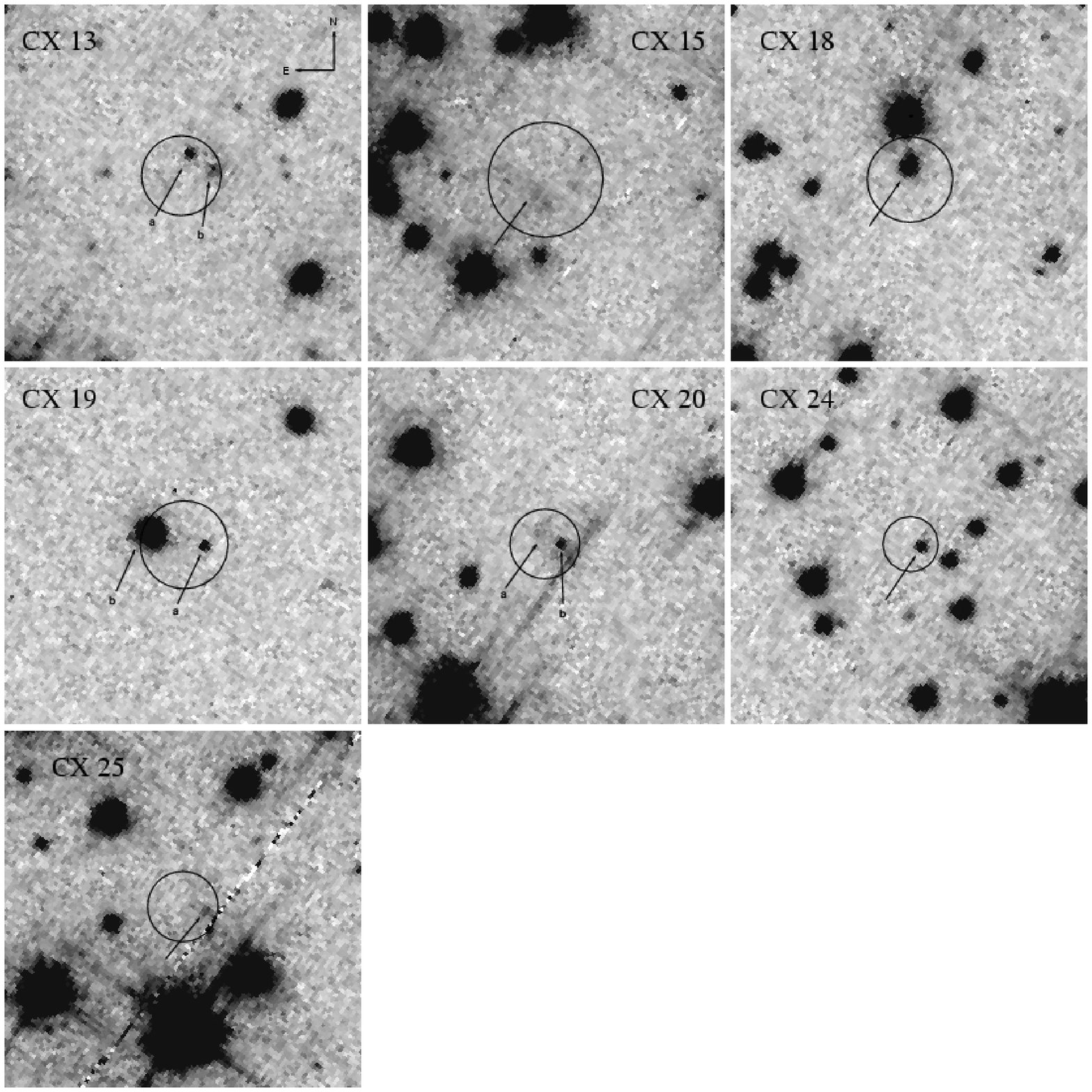,width=7.5in}
\caption{$5''\times5''$ finding charts for the candidate optical
  counterparts, obtained from \hst\ ACS observations. These images
  were taken in $B_{435}$. We have overlaid the 95\% error circles
  for the \chandra\ source positions and the candidate stars
  themselves are indicated by arrows. All images have the same grey
  scale.
}
\end{figure*} 
\clearpage

Lastly, we consider all the remaining \chandra\ sources outside the
half-mass radius; we do not possess useful optical data for any of
these.
Statistically we do not expect any of the sources outside the
 half-mass radius to be associated with NGC\,288 (see Section 2.2). 
From the \chandra\ deep field data, for the entire ACIS-S3 chip field of view, we expected
that 0--10 sources are associated with NGC 288. If
CX 18 and CX 24 (or indeed additionally CX 13,  CX 19 and CX 20) are CVs or chromospherically active
binaries (hence cluster members), then there are at most eight (five)
additional 
X-ray sources associated with NGC 288. 
In the absence of information of the optical colors of the counterparts,
 it is not possible to pursue this question further for individual cases.
 Nevertheless, we here describe the X-ray properties, {\em assuming
  cluster membership}. The X-ray fluxes of all \chandra\ sources are listed in Table 1
and 2, and the luminosities (assuming 8.4 kpc) range from $7\times10^{30}$ to $8\times10^{32}$\lum. CX 3 is one of the
three brightest X-ray sources. It has a relatively soft spectrum (see
Table 2) and is therefore inconsistent with a CV. It could be either a qLMXB or,
of course, yet another background galaxy. In contrast, the brightest source, CX 27, has a hard spectrum, favouring a CV identification; it is also noteworthy as an X-ray variable; the
\chandra\ spectrum indicates that it has a 0.3--7 keV flux of
$8-9\times10^{-14}$\flux, but the source was not detected with \rosat\
with a detection limit of $6\times10^{-14}$\flux. 
 Excluding
the brightest three sources (CX 3, CX 24 and CX 27), the remainder have
luminosities $< 10^{32}$\lum\ with an average of
$3\times10^{31}$\lum. This luminosity is at the lower end of quiescent
neutron stars in the field. Moreover, the X-ray colors are harder
than for neutron star systems. Hence, the remaining lower-luminosity X-ray sources outside the half-mass radius are very unlikely to be quiescent
neutron stars. Of these, the X-ray luminosities of CX 6, and CX 17 are too high
for any chromospherically active binary, and their relatively hard X-ray colors would indicate that they
are probably CVs, or absorbed AGN.


\begin{table*}
\centering{\small
    \caption{Scaling Parameters of M\,4, NGC\,6397,
    47~Tuc and NGC\,288\label{tab:scaling}}
\begin{tabular}{lcccccc}
    \hline \hline
      Cluster &
      log $\rho_0$ &
      $r_\mathrm{c}$ &
      $d$ &
      $M_V$ &
      $\Gamma$ &
      $M_\mathrm{h}$ \\
      &
      ($L_\sun\ \mathrm{pc}^{-3}$) &
      (\arcsec) &
      (kpc) &&&\\
\hline
    M\,4        & 4.01 & 49.8 & 1.73 & $-$6.9 & 1.0 & 1.0 \\
    NGC\,6397 & 5.68 &  3.0 & 2.3 & $-$6.6 & 2.1 & 0.76 \\
    47\,Tuc    & 4.81 & 24.0 & 4.5 & $-$9.4 & 24.9 & 10 \\
    NGC\,288   & 1.80 & 85.0 & 8.4 & $-$6.7 & 0.03 & 0.83\\
\hline
\end{tabular}
}
\par
\medskip
\begin{minipage}{0.85\linewidth}
Values for central density ($\rho_0$), core-radius
      ($r_\mathrm{c}$), distance ($d$) and absolute visual magnitude
      ($M_V$) originate from Harris\,1996 (version of February
      2003). For M4, the values of $\rho_0$ and $M_V$ are computed for
      the distance and reddening of Richer et al.\,(1997). The
      collision number is computed from $\Gamma \propto \rho_0^{1.5}\
      r_\mathrm{c}^2$ and the half-mass from $M_\mathrm{h} \propto
      10^{-0.4M_V}$. Values for $\Gamma$ and $M_\mathrm{h}$
      are normalized to the value of M4.
\end{minipage}
 \end{table*} 

\section{Discussion}

The luminosities of both good candidate members, CX24 and CX18, as
     well as those of the three less secure members CX13, CX19 and CX20,
are above
the lower limit of $4\times10^{30}$ ergs s$^{-1}$ in the 0.5 - 6.0 keV range
used in the study by Pooley et al.\ (2003) into the relation between
the stellar encounter rate and the incidence of X-ray sources in
globular clusters. To see whether NGC\,288 fits this relation we
compare its collision number $\Gamma\equiv{\rho_o}^{1.5}{r_c}^2$
(Verbunt 2003) with those of some other clusters, using the parameters
listed in Table 4. Here $\rho_o $ is the central density of the
cluster, and $r_c$ the core radius. The encounter number for NGC\,288
is a about 650 times smaller than that of 47\,Tuc, and 30 times
smaller than that of M\,4.  Pooley et al.\ (2003) reports 41$\pm$2
sources above the lower luminosity limit in 47\,Tuc (see also Grindlay
et al.\ 2001a; the uncertainty is due to the estimated number of
background sources), and thus, if the number of sources scales with
the encounter rate, the presence of two to five sources in NGC\,288 is a
very significant overabundance, even if we take into account small
number errors due to Poissonian fluctuations. The same conclusion is
reached on the basis of comparison with M\,4. This indicates that
the sources in NGC\,288 are not formed via stellar encounters.

Indeed, for magnetically active binaries, an origin from
a primordial binary is much more likely (Verbunt 2002). A scaling
with the total mass of the cluster is expected in this case, provided
that no large numbers of binaries have been destroyed by close encounters.
In a low-density cluster like NGC\,288, no such large scale destruction
has taken place. Since our information on low-luminosity X-ray sources
in most clusters is limited to the region within the half-mass radius,
we compare also the masses within this radius. By definition, however,
these masses are half of the total mass, and thus the scaling between
clusters is the same as for the total mass. If the visual
mass-to-light ratio is the same for all clusters listed in Table\,4,
the half masses scale with $10^{-0.4M_V}$. Thus the half-mass radii
of NGC\,6397 and NGC\,288 contain about 20\%\ less mass than the half-mass
radius of M\,4, which in turn contains a factor 10 less mass than the
half-mass radius of 47\,Tuc. Scaled by mass, the predicted number of 
active binaries $L_{0.5-6{\mathrm{keV}}}>4\times10^{30}$ erg\,s$^{-1}$
in NGC\,288 should be similar to those in NGC\,6397 and
M\,4, and about one tenth of those in 47\,Tuc. This is indeed observed.

We consider this strong evidence that magnetically active binaries in
globular clusters evolve from primordial binaries, much strengthening
the conclusion based by Bassa et al.\ (2004) on the source numbers in
M\,4.

As argued by Verbunt (2002), CVs take a position
in between low-mass X-ray binaries with a neutron star or black hole,
which are certainly formed from close encounters, and the magnetically
active binaries, formed from primordial binaries. The scaling of
source number with encounter number found for the sources with
$L_{0.5-6{\mathrm{keV}}}>4\times10^{30}$ ergs s$^{-1}$ by Pooley et al.\
(2003; see also Heinke et al.\ 2003b) suggests that CVs are mostly
made via stellar encounters as well. This is in
agreement with the result by Davies (1997) that the formation of
CVs via evolution from -- relatively wide --
primordial binaries is suppressed by the destruction of such binaries
in dense cores of globular clusters. We do not expect
even a single CV in NGC\,288 if we scale with encounter
numbers from any of the other clusters listed in Table\,4. However,
neither the observed scaling law nor the theoretical computations extends to clusters with core densities or encounter rates as low as that of NGC\,288.
According to the computations by Davies (1997) a cluster core with a star
density of $1000$ pc$^{-3}$ allows most of the CV
progenitors to evolve into a CV.  It is
therefore probable that the CV, CX 24, in
cluster with an even lower central number density, evolved
from a primordial binary. As an even more extreme case than NGC\,288,
recent \chandra\ observation of the old open cluster M\,67 
reveals a large number of BY Dra and RS CVn systems as well as interacting
binary candidates (van den Berg et al. 2004). It is interesting to
note that the total X-ray luminosity of M\,67 is dominated by binaries
with giants, whereas no such binaries have been securely identified as
optical counterpart to an X-ray source in any globular cluster so far.
Investigations of other globular clusters
with low density cores must be done to verify these conclusions. 

\vspace{2mm}
\begin{inlinefigure}
\psfig{file=f6.ps,width=3.3in}
\caption{X-ray luminosity as a function of the absolute magnitude, for
low-luminosity X-ray sources in globular clusters. The large symbols
in this Figure indicate the optically identified X-ray sources in the
field of view of our Chandra observation of NGC\,288, {\em where we
compute absolute magnitude and X-ray luminosity under the assumption
that the sources are cluster members}. This assumption is probably
correct for the candidate CV (large triangle) and
the three candidate active binaries (large stars); the sources indicated
with a large square are probably extragalactic sources. The smaller
symbols in this Figure indicate objects found in other clusters, mostly
47\,Tuc and M\,4 (see Figure 6 of Bassa et al.\ 2004). Since we do not have
obsrvations of NGC\,288 in the $V$-band, we estimate $V$ from
$V=0.5(B_{435}+r_{625})$. The dashed line of constant X-ray to optical
flux ratio roughly separates CVs from active binaries.}
\end{inlinefigure}

\begin{acknowledgements}
We thank Andrew Dolphin for providing a modified version of his
DOLPHOT code.
Support for Proposal number 10120 and 9959 was provided by NASA through a grant from the Space Telescope Science Institute, which is operated by the Association of Universities for Research in Astronomy, Incorporated, under NASA contract NAS5-26555. 
\end{acknowledgements}

\end{document}